\documentclass[chi_draft]{sigchi}

\CopyrightYear{2016}

\toappear{
}

\usepackage{balance}       
\usepackage{graphics}      
\usepackage[T1]{fontenc}   
\usepackage{txfonts}
\usepackage{mathptmx}
\usepackage[pdflang={en-US},pdftex]{hyperref}
\usepackage{color}
\usepackage{booktabs}
\usepackage{textcomp}

\usepackage{microtype}        
\usepackage{ccicons}          
\usepackage{subcaption}

\def\plaintitle{Exploring Configurations for Multi-user Communication in Virtual Reality}

\def\emptyauthor{}
\def\plainkeywords{Multi-user communication; computer-supported collaborative work; shared workspace}

\makeatletter
\def\url@leostyle{%
  \@ifundefined{selectfont}{
    \def\UrlFont{\sf}
  }{
    \def\UrlFont{\small\bf\ttfamily}
  }}
\makeatother
\def\pprw{8.5in}
\def\pprh{11in}

\definecolor{linkColor}{RGB}{6,125,233}
\hypersetup{%
  pdftitle={\plaintitle},
  pdfauthor={\emptyauthor},
  pdfkeywords={\plainkeywords},
  pdfdisplaydoctitle=true, 
  bookmarksnumbered,
  pdfstartview={FitH},
  colorlinks,
  citecolor=black,
  filecolor=black,
  linkcolor=black,
  urlcolor=linkColor,
  breaklinks=true,
  hypertexnames=false
}

\begin{document}

\title{\plaintitle}

\numberofauthors{3}
\author{%
  \alignauthor{Zhenyi He\\
    \affaddr{New York University}\\
    \affaddr{New York, United States}\\
    \email{zh719@nyu.edu}}\\
  \alignauthor{Karl Rosenberg\\
    \affaddr{New York University}\\
    \affaddr{New York, United States}\\
    \email{ktr254@nyu.edu}}\\
  \alignauthor{Ken Perlin\\
    \affaddr{New York University}\\
    \affaddr{New York, United States}\\
    \email{ken.perlin@gmail.com}}\\
}

\maketitle

\begin{figure*}
\centering
    \begin{subfigure}[b]{0.6\columnwidth}
        \includegraphics[width=1\columnwidth]{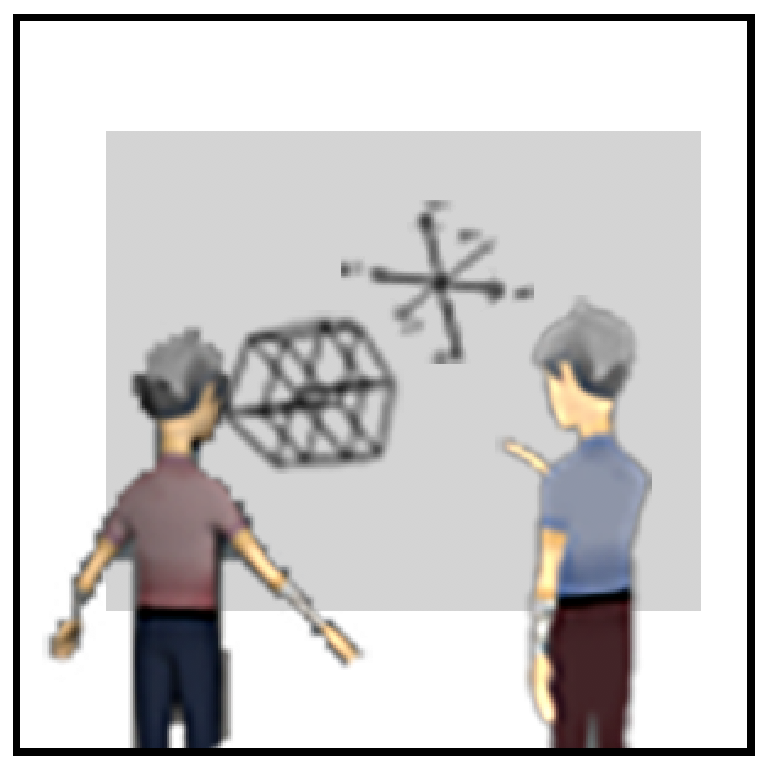}
        \caption{Side-by-side Configuration}
    \end{subfigure}
    \begin{subfigure}[b]{0.6\columnwidth}
        \includegraphics[width=1\columnwidth]{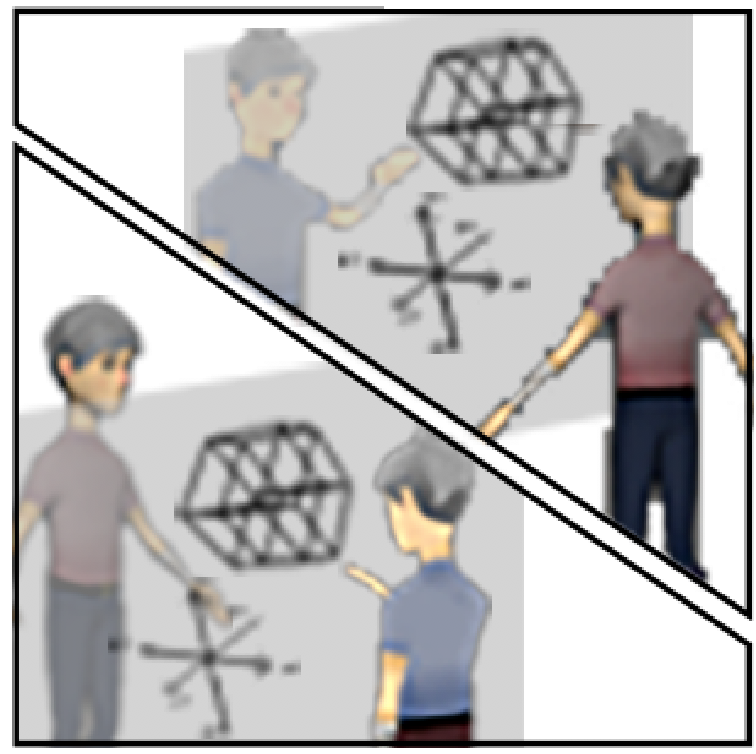}
        \caption{Mirrored Face-to-face Configuration}
    \end{subfigure}
    \begin{subfigure}[b]{0.6\columnwidth}
        \includegraphics[width=1\columnwidth]{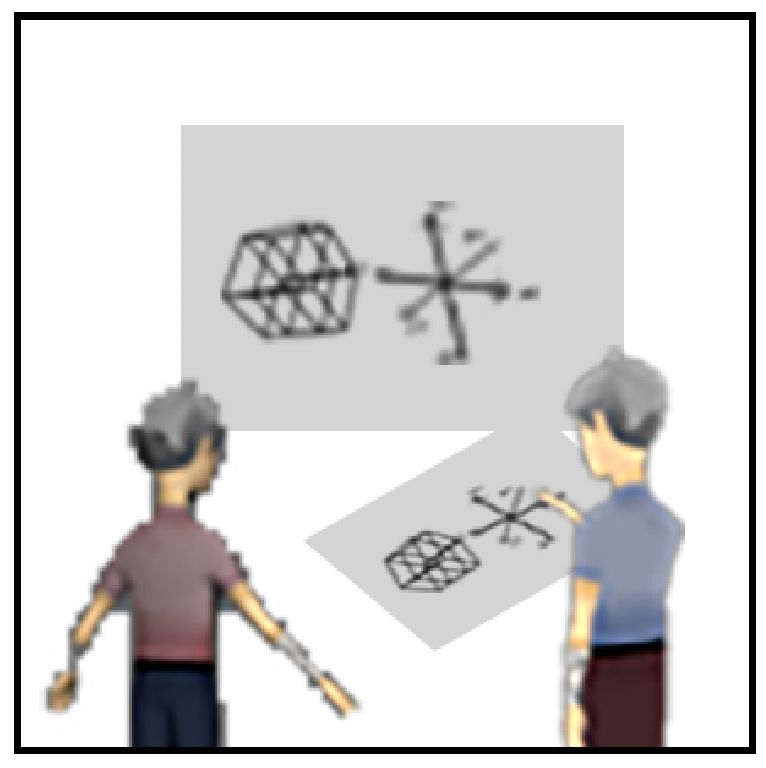}
        \caption{Eyes-free Configuration}
    \end{subfigure}
  \caption{Side view of multiple configurations}~\label{fig:teaser}
\end{figure*}

\begin{abstract}
  Virtual Reality (VR) enables users to collaborate while exploring scenarios not realizable in the physical world. We propose CollabVR, a distributed multi-user collaboration environment, to explore how digital content improves expression and understanding of ideas among groups. To achieve this, we designed and examined three possible configurations for participants and shared manipulable objects. In configuration (1), participants stand side-by-side. In (2), participants are positioned across from each other, mirrored face-to-face. In (3), called "eyes-free," participants stand side-by-side looking at a shared display, and draw upon a horizontal surface. We also explored a "telepathy" mode, in which participants could see from each other's point of view. We implemented "3DSketch" visual objects for participants to manipulate and move between virtual content boards in the environment. To evaluate the system, we conducted a study in which four people at a time used each of the three configurations to cooperate and communicate ideas with each other. We have provided experimental results and interview responses.

\end{abstract}

\category{H.5.3.}{Information Interfaces and Presentation}{Computer-supported cooperative work}
\category{H.5.3.}{Information Interfaces and Presentation}{Miscellaneous}


\section{Introduction}
Virtual Reality (VR) is being explored increasingly, spurred by the availability of high quality consumer headsets in recent years. VR enables rich design spaces in HCI by providing 3D input and immersive experiences. In 1998, an idea, the "Office of the future," was proposed to allow remotely located people to feel as though they were together in a shared office space~\cite{raskar1998office}, via a hybrid of modalities including telepresence, large panoramic displays and manipulation of shared 3D objects. The core idea was that VR had the potential to enhance communication and collaboration among groups of people. Since then, significant progress has been made in exploring techniques for communication~\cite{ishii1993integration, otsuka2016mmspace}, collaborative works~\cite{tang2010three, kunert2014photoportals}, infrastructure~\cite{o2011blended, thomas2014muvr} and various modalities~\cite{follmer2013inform} for multi-user experiences. 
For example, SynchronizAR designed a registration algorithm for mobile AR so the participants could join the co-located experience and not need to take extra steps to ensure good-quality positional tracking~\cite{huo2018synchronizar}. SpaceTime focused on improving the experience for two experts collaborating on design work together~\cite{xia2018spacetime}. InForce created a set of novel interaction techniques including haptic feedback for distributed collaboration~\cite{nakagaki2019inforce}. Many such collaboration systems allow users to communicate as in everyday life, without enhancement. Less studied is how VR can enhance communication and be integrated into various kinds of collaborative work.

In daily life while speaking to others, we commonly use gestures or visual aids~\cite{tversky2003human} to help present ideas, either subconsciously or purposefully. Visual aids can be drawn on paper, a whiteboard\cite{cherubini2007let}, or a screen via desktop sharing in video conferences.
A key task for collaborators is visually, physically, and cognitively following the content being drawn and the people who are speaking with. They need a clear view of the content under discussion and need to be aware of other collaborators' presence when communicating~\cite{ishii1992clearboard}. They should also understand and be able to reason about what is being shown and said.
According to Regenbrecht et al.~\cite{regenbrecht2015mutual}, facial interaction like eye contact and mutual gaze has always been recognized as an important requirement for effective visual communications. That suggests it is helpful during communication to look at the person who is talking or at least be aware of the person's location. 
Oftentimes, participants in a group discussion also need to move around physically to follow the content and each other. In addition, following and presenting content requires the ability to express ideas clearly as well as a fair understanding of the ideas. This may become particularly challenging in discussions revolving around, for example, the placement and design of 3D objects, which cannot easily be visualized using traditional 2D communication media (paper, whiteboard, and so on).

Prior work tried to improve communication using alternative arrangements of spaces and enhanced digital content for communication. One trend has been to help people maintain face-to-face interaction during communication. ClearBoard~\cite{ishii1993integration} created a shared workspace in which two users collaborate remotely without losing all the advantages of in-person face-to-face interactions. MMSpace changed participants' poses and positions automatically to mirror the remote users' heads to enable face-to-face interactions~\cite{otsuka2016mmspace}. Tan et al. built a face-to-face presentation system for remote audiences ~\cite{gazeAwareness}. Interacting with digital content in a shared space is common too. Three's Company~\cite{tang2010three} explored collaborative activities over connected digital tabletops which support a shared sense of presence among people and artifacts associated with the task. Tele-Board~\cite{gumienny2011tele} designed a groupware system specialized in creative working modes using traditional whiteboard and sticky notes in digital form for distributed users.

Still, it is unclear how the arrangement of users and content in a VR space and how advanced digital content affect communication for co-located and distributed collaboration. We implemented CollabVR, a multi-user VR collaboration environment, to explore how
spatial arrangements of people and visual content impact communication.
\begin{itemize}
\item Three different \textit{Configurations}: 1) side-by-side, 2) mirrored face-to-face and 3) eyes-free. A \textit{Configuration} is a spatial arrangement including the digital content and the participants. Each \textit{Configuration} explores a different arrangement for the shared workspace.
\item \textit{Telepathy mode}, which gives participants the ability to view the world from another user's perspective, either through a windowed overlay or immersively. We refer to the person who chooses to see from someone else's perspective as the "observer." The person whose perspective is chosen is the "observee."
\item \textit{3DSketch}, which is an enhanced Chalktalk sketch (see ~\ref{sec:ct}) which can be manipulated in the immersive environment. Rich interactions are designed so users can quickly share are demonstrate different ideas with each other.
\end{itemize}
We conducted a user study with groups of four people at a time. They were asked to experience each configuration and complete the same tasks for each of them. Afterwards, we proceeded to user interviews in which participants shared their thoughts on the three \textit{Configurations}, \textit{telepathy mode} and the usability of the CollabVR communication platform.

\section{Related Work}
Here we present previous work focusing on manipulation of interactive content, collaborative work for co-located and distributed groups, and enhancement of communication with immersive environments.

\subsection{Manipulation on Interactive Contents}
Some work made use of physical proxies to manipulate digital content directly. Photoportals allowed direct manipulation of 3D references to objects, places, and moments by moving the physical proxy directly~\cite{kunert2014photoportals}. MAI Painting Brush imitated a real paint brush, and constructed a mixed reality
(MR) painting system that enabled direct painting on physical
objects~\cite{otsuki2010mai}. Eckard et al. designed a multi-modal actuated Tangible User Interface(TUI) for distributed collaboration, so users could control the remote items by moving the local corresponding items~\cite{riedenklau2012integrated}. Physical Telepresence proposed a series of shared work spaces so remote physical objects and physical renderings of shared digital content could be manipulated locally~\cite{leithinger2014physical}. In our system, we also seek to make direct manipulation of interactive content straightforward and intuitive. Similarly, the interactions we designed for \textit{3DSketch} ~\ref{3dsketch} follow that philosophy as well.

\subsection{Co-located and Distributed Collaboration}
Much work has been done in collaborative applications in VR/MR. T(ether) is a spatially-aware display system for co-located collaborative manipulation and animation of objects ~\cite{lakatos2014t}. Trackable markers on pads and digital gloves allow participants to use gestures to manipulate objects in space. Virtual Replicas for Remote Assistance is a remote collaboration system, allowing a remote expert to guide local users to assemble machine parts by using virtual replicas~\cite{oda2015virtual}.
Martin et al. proposed an MR system that allowed distributed users to align and understand each other's perspective by video feed~\cite{feick2018mixed}.
SpaceTime is a scene editing tool supporting multi-user collaboration in VR, either co-located or remote~\cite{xia2018spacetime}. To support conflict resolution when two users wish to work on the same object simultaneously, SpaceTime creates per-user branches of the object, which can later be automatically merged or manually resolved. Most of the system supported audio for general communication. Our system aims to enhance the experience of general communication, which can benefit various kinds of more specific collaborative work.

\subsection{Communication in VR/MR}
Some previous work contributed to rendering remote participants to give them presence during communication. Hrvoje Benko et al. proposed a unique spatial AR system that combines dynamic projection mapping and multiple perspective views to support face-to-face interaction~\cite{benko2014dyadic}. MetaSpace did full body tracking for distributed users to create a better sense of presence~\cite{sra2015metaspace}. Holoportation demonstrated 3D reconstructions of an entire space, including people~\cite{orts2016holoportation}. 

Some previous work contributed to communication in immersive environments. ClearBoard allowed a pair of users to shift easily between interpersonal space and a shared workspace~\cite{ishii1993integration}.
The key metaphor of ClearBoard is ``talking through and drawing on a big transparent glass board.'' No gaze or eye contact information is lost while working on the content. ShareVR enables communication between an HMD user and a non-HMD user~\cite{gugenheimer2017sharevr}. By using floor projection and mobile displays to visualize the virtual world, the non-HMD user is able to interact with the HMD user and become part of the VR experience. The work discusses how people with different devices communicate with each other. MMSpace allows for face-to-face social interactions and telepresence in the context of small group remote conferences~\cite{otsuka2016mmspace}. It uses custom-built mechanical displays on which images of remote participants are projected, and which move in response to users' movements. Pairs of participants can maintain eye contact with each other and remain aware of each other's focus. 
TwinSpace supports deep interconnectivity and flexible mappings between virtual and physical spaces~\cite{reilly2010twinspace}.
Your Place and Mine explored three ways of mapping two differently sized
physical spaces to shared virtual spaces to understand how social presence, togetherness, and movement are influenced~\cite{sra2018}.



Some previous works aiming to improve communication with VR experimented with the placement of users and content, but did not focus on looking at how these factors affected users' ability to follow content and other people during communication. We designed and implemented three different \textit{configurations} as well as \textit{telepathy mode} to investigate how communication could be improved with VR, using \textit{3DSketch}. In the following section, we discuss the design of CollabVR and its \textit{configurations}.

\section{CollabVR Design}
\subsection{Configurations}

\begin{figure}
    \centering
    \begin{subfigure}[b]{0.3\columnwidth}
        \includegraphics[width=1\columnwidth]{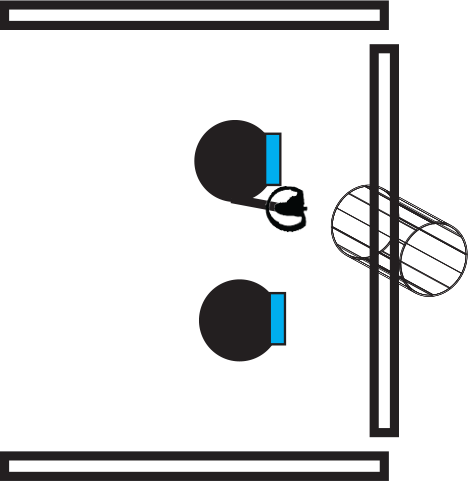}
        \caption{two people in side-by-side configuration}
        \label{fig:sbs2}
    \end{subfigure}
    ~ 
    \begin{subfigure}[b]{0.3\columnwidth}
        \includegraphics[width=1\columnwidth]{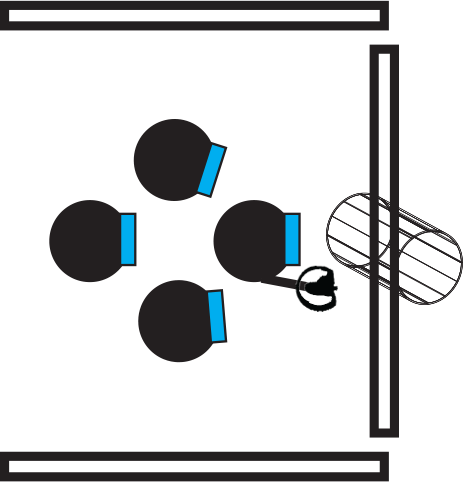}
        \caption{four people in side-by-side configuration}
        \label{fig:sbs4}
    \end{subfigure}
    ~ 
    \begin{subfigure}[b]{0.3\columnwidth}
        \includegraphics[width=1\columnwidth]{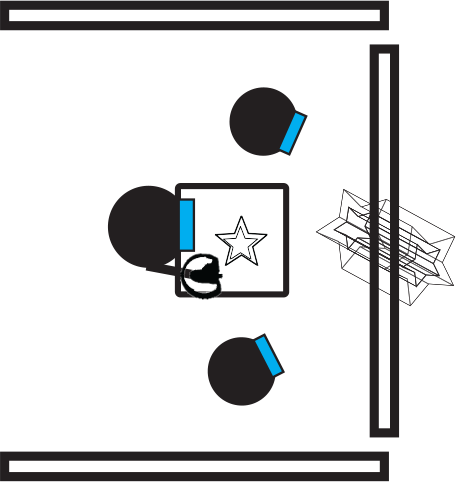}
        \caption{people in eyes-free configuration}
        \label{fig:ef}
    \end{subfigure}
    \begin{subfigure}[b]{0.3\columnwidth}
        \includegraphics[width=1\columnwidth]{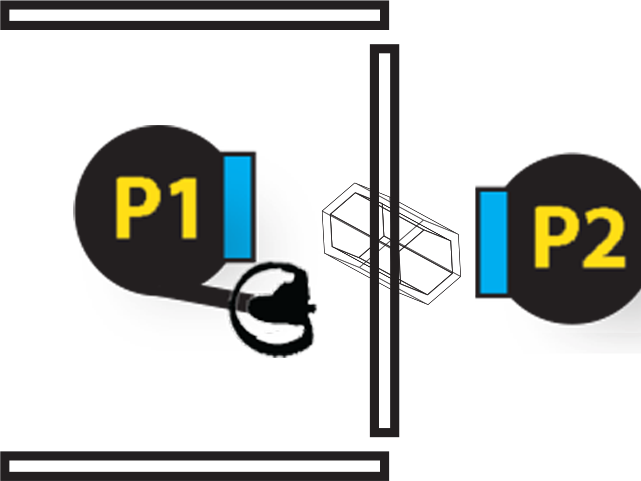}
        \caption{two people in mirrored face-to-face configuration in P1's view}
        \label{fig:f2f1}
    \end{subfigure}
    ~ 
    \begin{subfigure}[b]{0.3\columnwidth}
        \includegraphics[width=1\columnwidth]{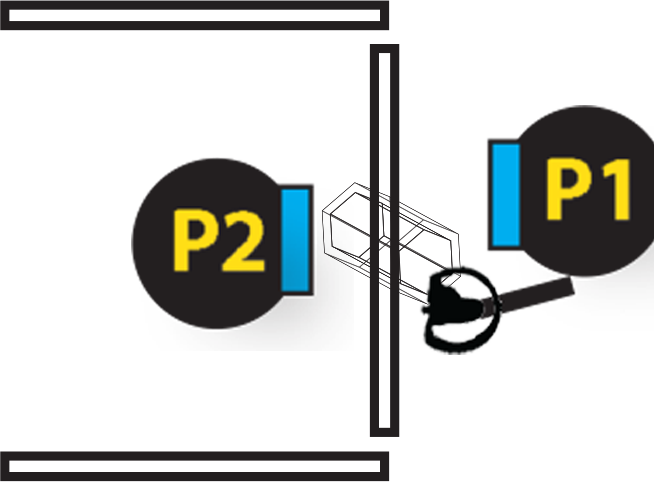}
        \caption{two people in mirrored face-to-face configuration in P2's view}
        \label{fig:f2f3}
    \end{subfigure}
    ~ 
    \begin{subfigure}[b]{0.3\columnwidth}
        \includegraphics[width=1\columnwidth]{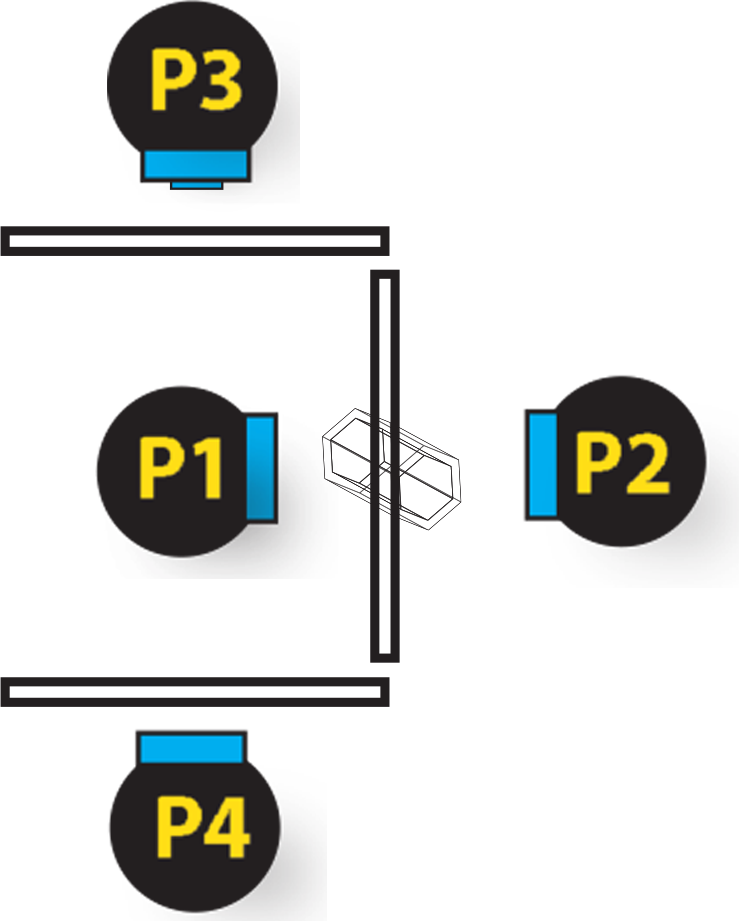}
        \caption{four people in mirrored face-to-face configuration}
        \label{fig:f2f4}
    \end{subfigure}
    \caption{Three Configurations.}\label{fig:config}
\end{figure}
\textit{Configurations} refer to the placement of participants and digital content in the immersive environment. The three \textit{Configurations} are inspired by daily experience and previous work, see figure ~\ref{fig:teaser}. Users can switch between them freely.
\subsubsection{Side-by-side}
The \textit{Side-by-side Configuration} is based on whiteboard brainstorming scenarios. In this configuration, all the participants have the same vision on the contents and the other participants.


\subsubsection{Mirrored Face-to-face} \label{mf2f}
The \textit{Mirrored Face-to-face Configuration} comes from daily communication. We are used to looking at each other to receive feedback from others and make gestures as part of our explanations during a discussion. For this configuration, the users are face-to-face in the virtual environment. The interactive content board is placed between them, so each user sees the others on the opposite side of the content board, left-right reversed as if reflected in a mirror. The challenge is ensuring all the content is consistent for everyone on each side of the board. Mirror reversal allows, for example, text to be readable and asymmetric objects to appear correct for each participant.
ClearBoard~\cite{ishii1993integration} implemented "mirror reversal" via video capture and projection techniques to solve a similar problem for 2D displays. Inspired by that, we implemented a 3D immersive mirror reversal for our MR configuration. We place the users physically on the same side of the content board and mirror all other users (from one user's perspective) to the other side. This way, information such as gaze and gesture direction is preserved, so participants can know where each other is looking and pointing. When multiple content-boards are active, an individual user will see the other users mirror reversed over the specific content board they are looking at. Users have face-to-face interactions in this multiple board scenario if they look at the same board (see figure~\ref{fig:f2f4}). 

\subsubsection{Eyes-free}
The \textit{Eyes-free Configuration} provides two duplicate boards where in the other configurations there would be one board. One board is vertical like a whiteboard and the other is horizontal like the surface of a table. The horizontal surface appears only for the person working at a particular board and can be scaled to allow the user to draw with varying precision. 
The design allows users to draw on a horizontal surface, which leads to less fatigue over the course of long-term activity. We render the drawings on the vertical board for all users to see together, as well as on the smaller horizontal board. There is potential for less fatigue because a user can draw with the hand on the scaled-down horizontal board over a shorter distance than would be required with the other configurations. Unlike in the other configurations, which have the user draw on a vertical board, it is unnecessary to move the whole arm to draw. In this \textit{Configuration}, users are free to look down at the surface as if they were drawing on a piece of paper. Alternatively, they may choose to look straight forward at the vertical board while drawing on the horizontal board using forward/backward movements mapped to up/down movements, as if using a mouse and computer display. Content on the horizontal drawing board is projected down to be flat so 3D content does not obscure the view of the vertical board, but the same content is simultaneously displayed in 3D on the vertical board for viewing of the final result of the drawing. 
\subsection{Telepathy Mode}

\begin{figure}
    \centering
    \begin{subfigure}[b]{0.4\columnwidth}
        \includegraphics[width=1\columnwidth]{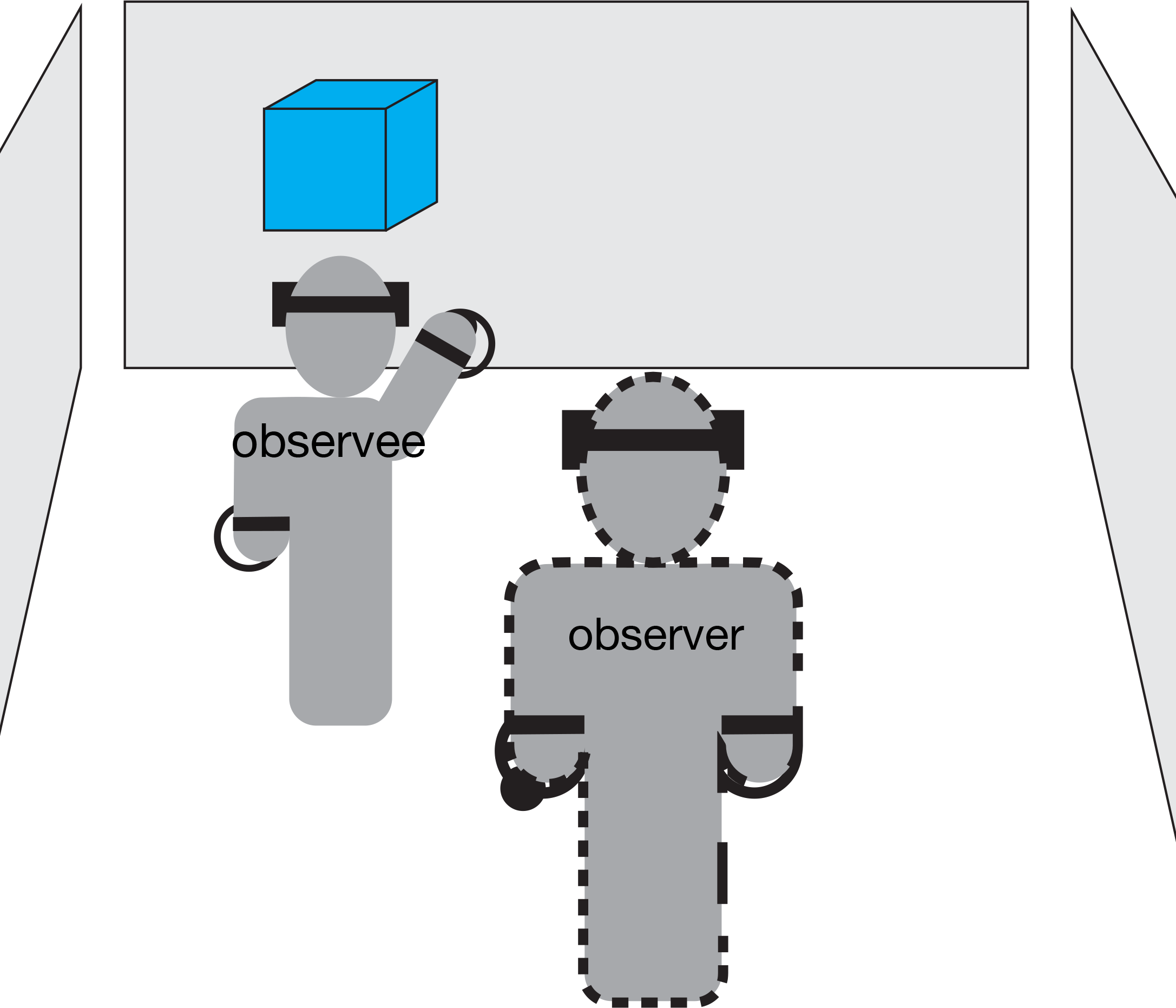}
        \caption{Immersive Telepathy}
        \label{fig:observeI}
    \end{subfigure}
    ~ 
    \begin{subfigure}[b]{0.4\columnwidth}
        \includegraphics[width=1\columnwidth]{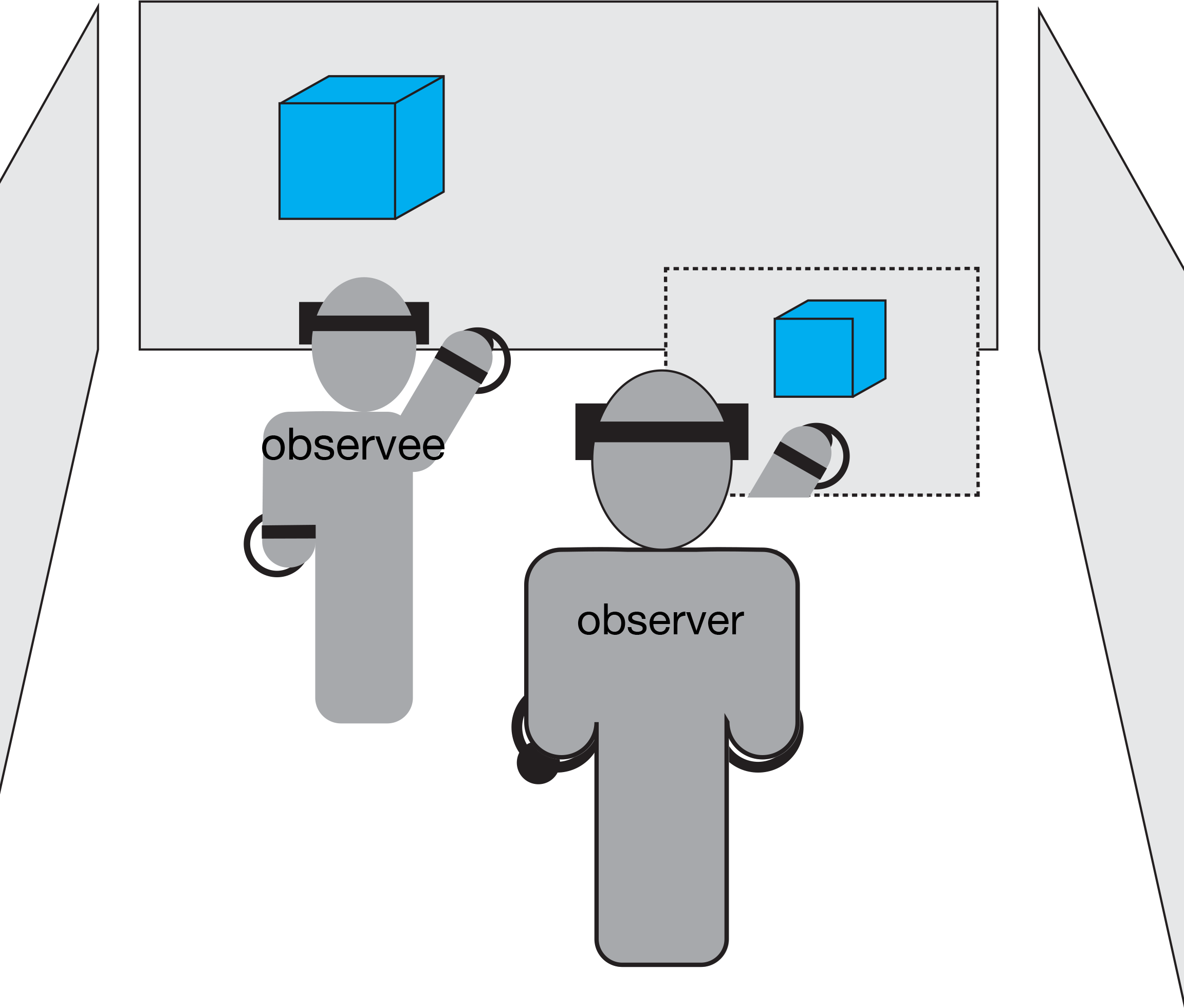}
        \caption{Windowed Telepathy}
        \label{fig:observeW}
    \end{subfigure}
    ~ 
    \begin{subfigure}[b]{0.4\columnwidth}
        \includegraphics[width=1\columnwidth]{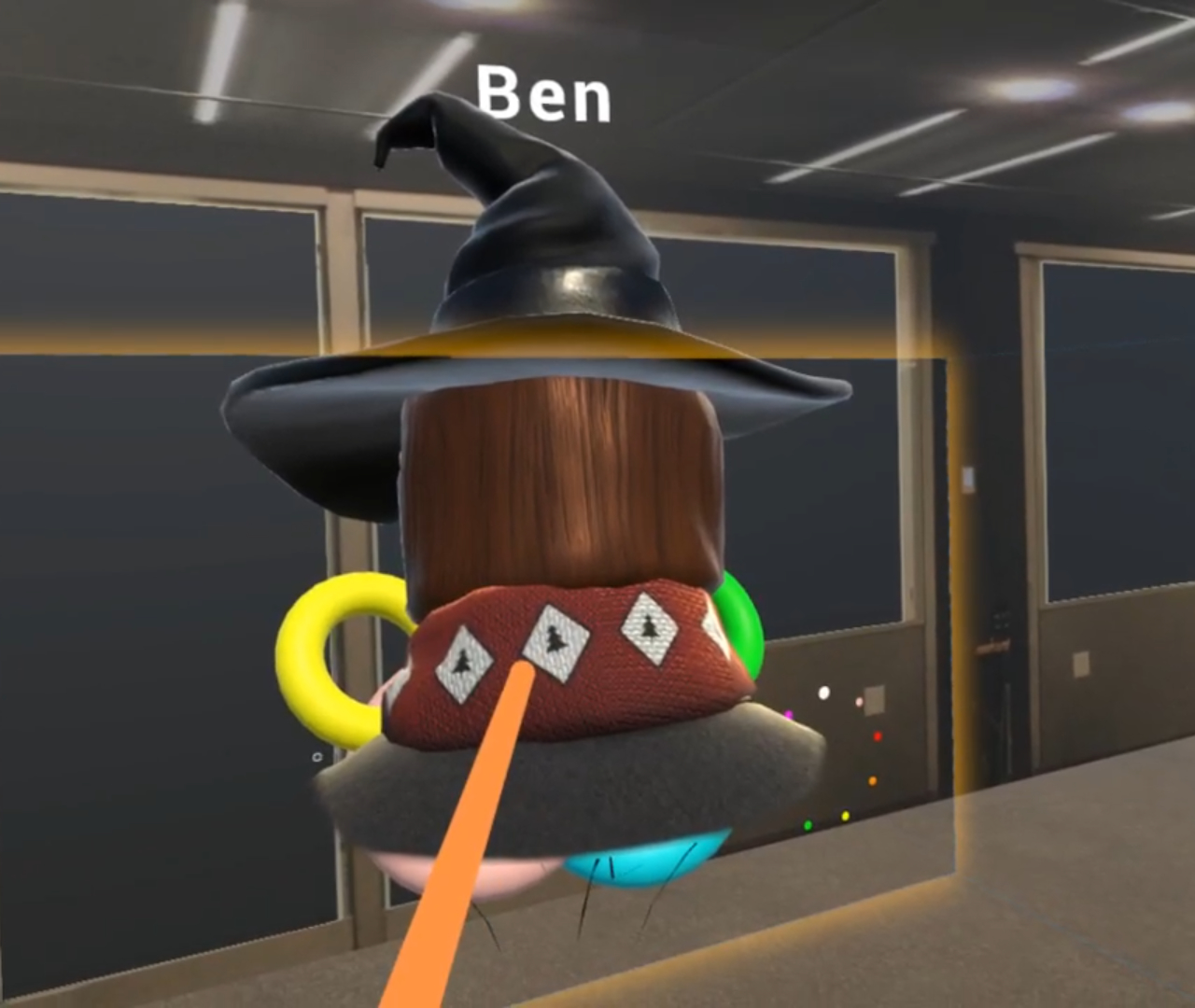}
        \caption{Choosing the Observee}
        \label{fig:observe1}
    \end{subfigure}
    \begin{subfigure}[b]{0.4\columnwidth}
        \includegraphics[width=1\columnwidth]{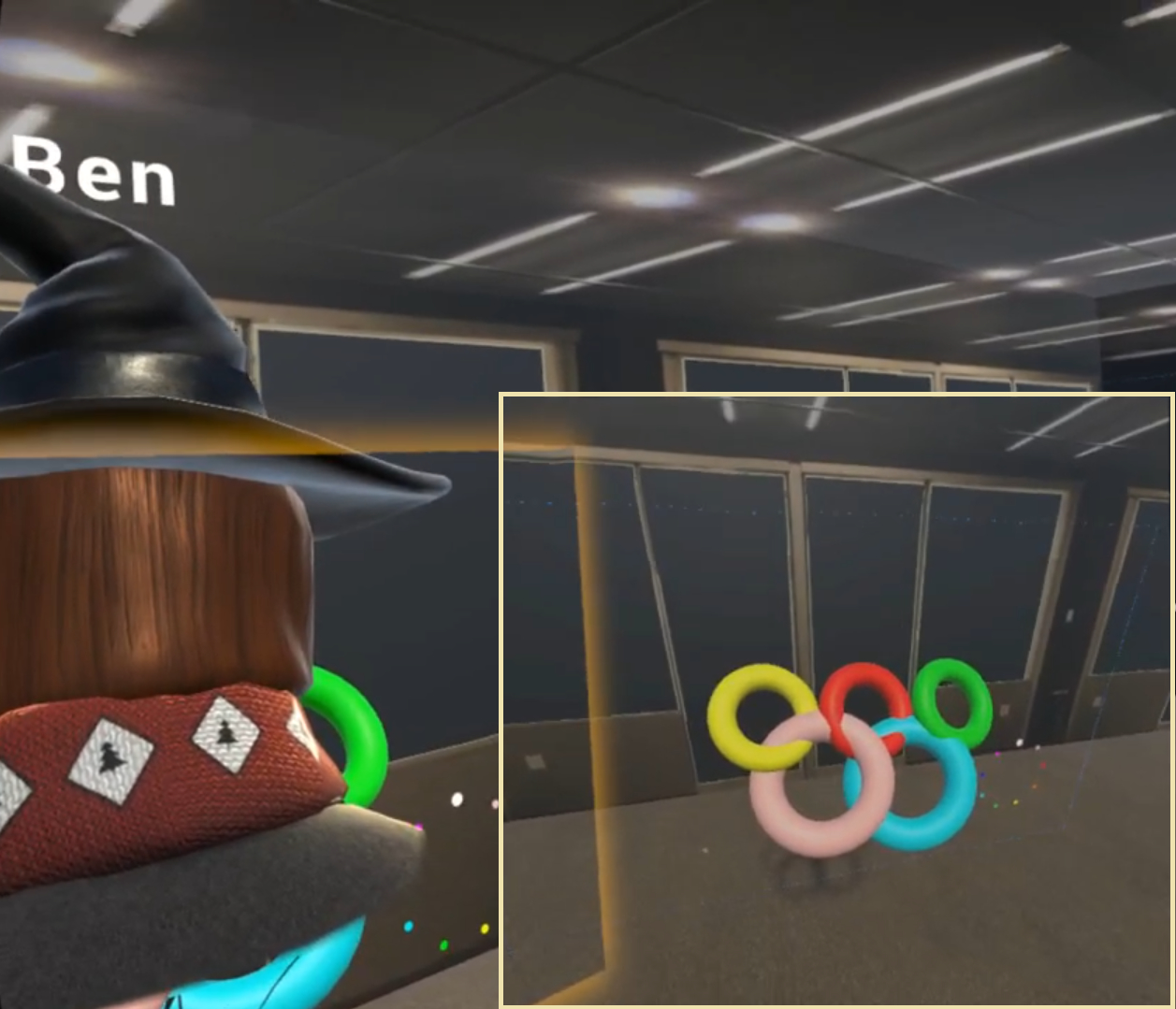}
        \caption{Observing the Observee}
        \label{fig:observe2}
    \end{subfigure}
    \caption{telepathy mode. (a) Shows how immersive telepathy works. After choosing the observee, the observer (the dotted outline avatar) will disappear from his/her original location in the world and move to the observee's position. (b) Shows how windowed telepathy looks. A semi-transparent telepathy window will be placed relative to where the observer is, so it is visible in the corner during telepathy. (c) and (d) show how it looks. In (d) we can see that the shapes are no longer blocked from view.}\label{fig:observe}
\end{figure}

We created an telepathy mode to give people the ability to see the environment from another person's point-of-view. Among other use cases, it is to help people follow the content and the person who is drawing, as well as see content that may be blocked from another angle, see figure~\ref{fig:observe1}.
\begin{itemize}
    \item Immersive Telepathy. The observer will be teleported to the observee's position and see the world from his/her viewpoint. At that time, the observer is free to look around, so the observer and observee can guide each other to look at whatever is of interest. We have first person perspective and third person perspective options for this alternative (see figure~\ref{fig:observeI}).
    \item Windowed Telepathy. The observer can see the observee's first-person view displayed on a small window overlay placed on the observer's periphery, (see figure~\ref{fig:observeW}).
\end{itemize}
\textit{telepathy mode} became more complex to implement for the \textit{Mirrored Face-to-face Configuration} and \textit{Eyes-free Configurations} than for the {Side-by-side Configuration}. Since we are not streaming rendering results from each participant to the rest, telepathy mode is implemented locally using other participants' position and rotation to place and orient the local representations of remote participants. For the \textit{Mirrored Face-to-face Configuration}, originally the observer (the local user) is the only participant who is not mirror reversed to the other side of the content board. The other participants, including the potential observee, will be mirror reversed over the content board they are working on. When \textit{telepathy mode} is on, the environment will be rendered in the observee's view, so the observee is then placed at the original side of the board in order to give the observer a correct view of the board from the observee's perspective. For the windowed telepathy mode, we need to change the positions and rotations for the main view and the observer view.

\subsection{3DSketch} \label{3dsketch}
The term \textit{3DSketch} comes from the \textit{Sketch} objects in Chalktalk~\cite{perlin2018chalktalk}, which will be briefly introduced in section~\ref{sec:ct}. A \textit{Sketch} is a combination of interactive graphical elements. We designed \textit{3DSketch} to provide three-dimensional manipulation of \textit{Sketches}. To enable quick demonstrations and accommodate changing ideas, we support necessary operations which are accessible via a pie menu that appears when a user selects a 3DSketch. Operations include copying, scaling, and rotation of the 3DSketch. To translate a 3DSketch, the user moves it directly with the hand. Users are free to copy the items on one board to another board and work on their own version. Also, each participant is able to work on different parts of the 3DSketches and then integrate them into one digital object.

\begin{figure}
    \centering
    \begin{subfigure}[b]{0.4\columnwidth}
        \includegraphics[width=1\columnwidth]{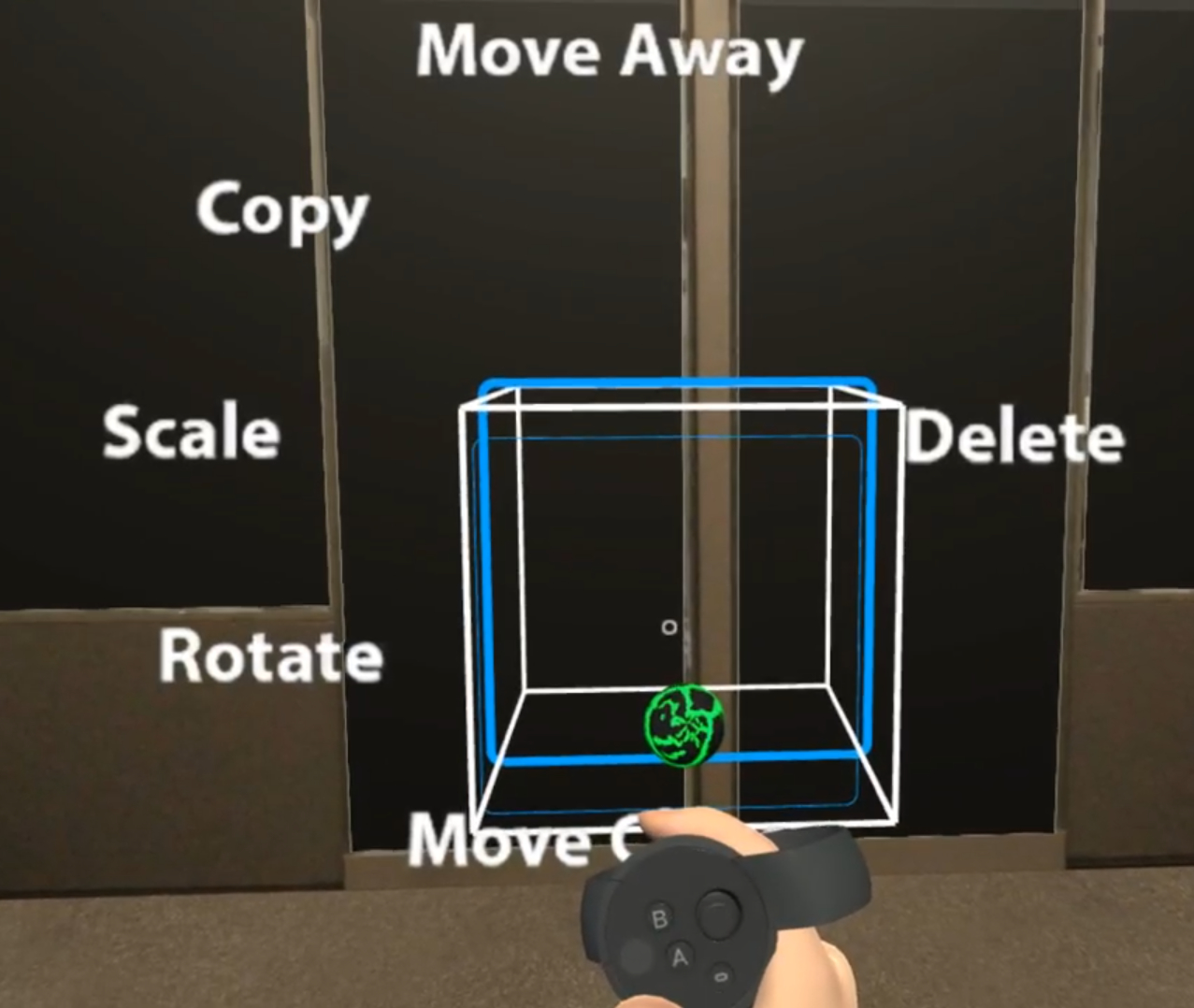}
        \caption{a cube 3DSketch has been selected and a pie menu is shown}
    \end{subfigure}
    \begin{subfigure}[b]{0.4\columnwidth}
        \includegraphics[width=1\columnwidth]{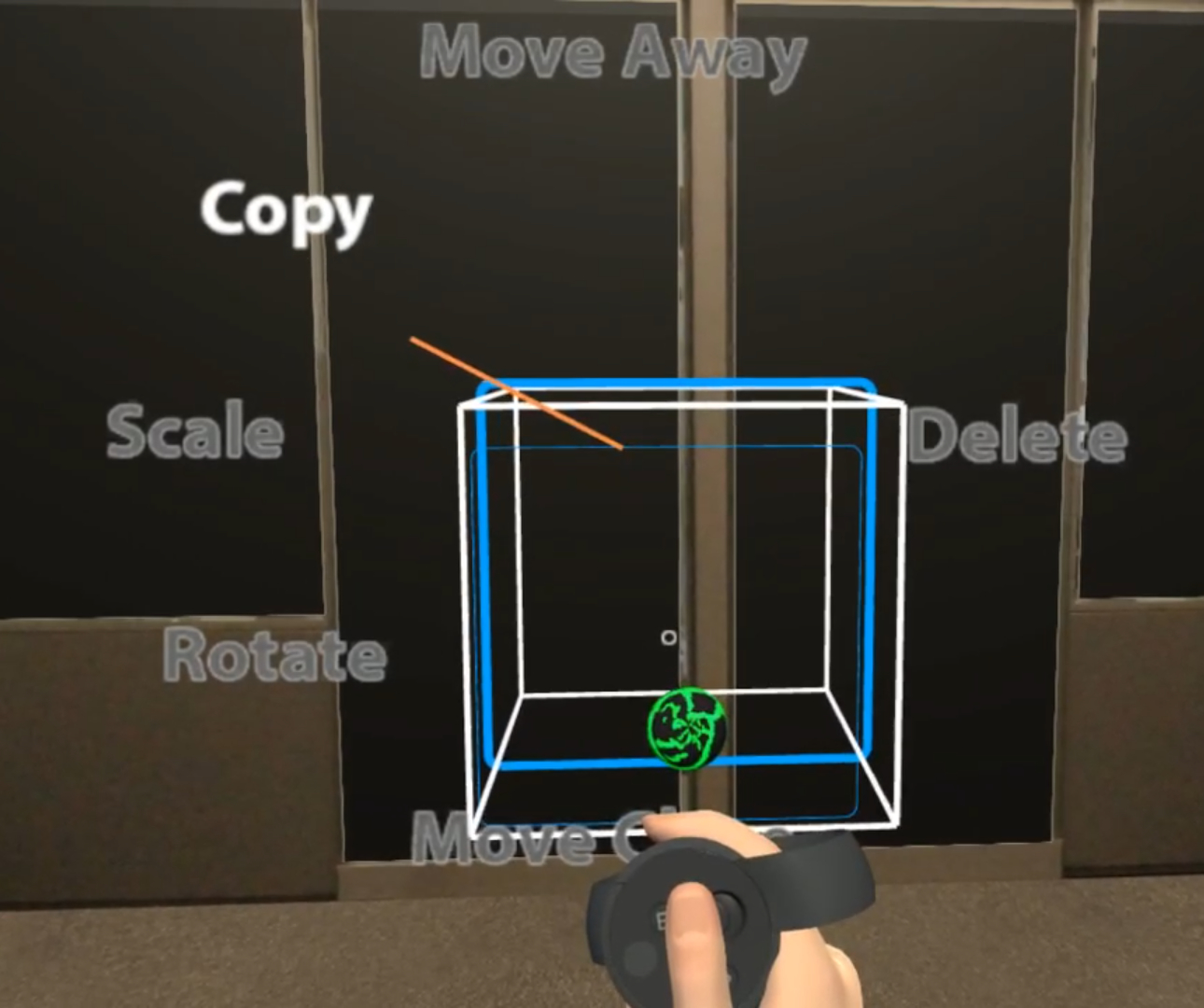}
        \caption{the user moves the piece menu cursor to the copy operation}
    \end{subfigure}
    \begin{subfigure}[b]{0.4\columnwidth}
        \includegraphics[width=1\columnwidth]{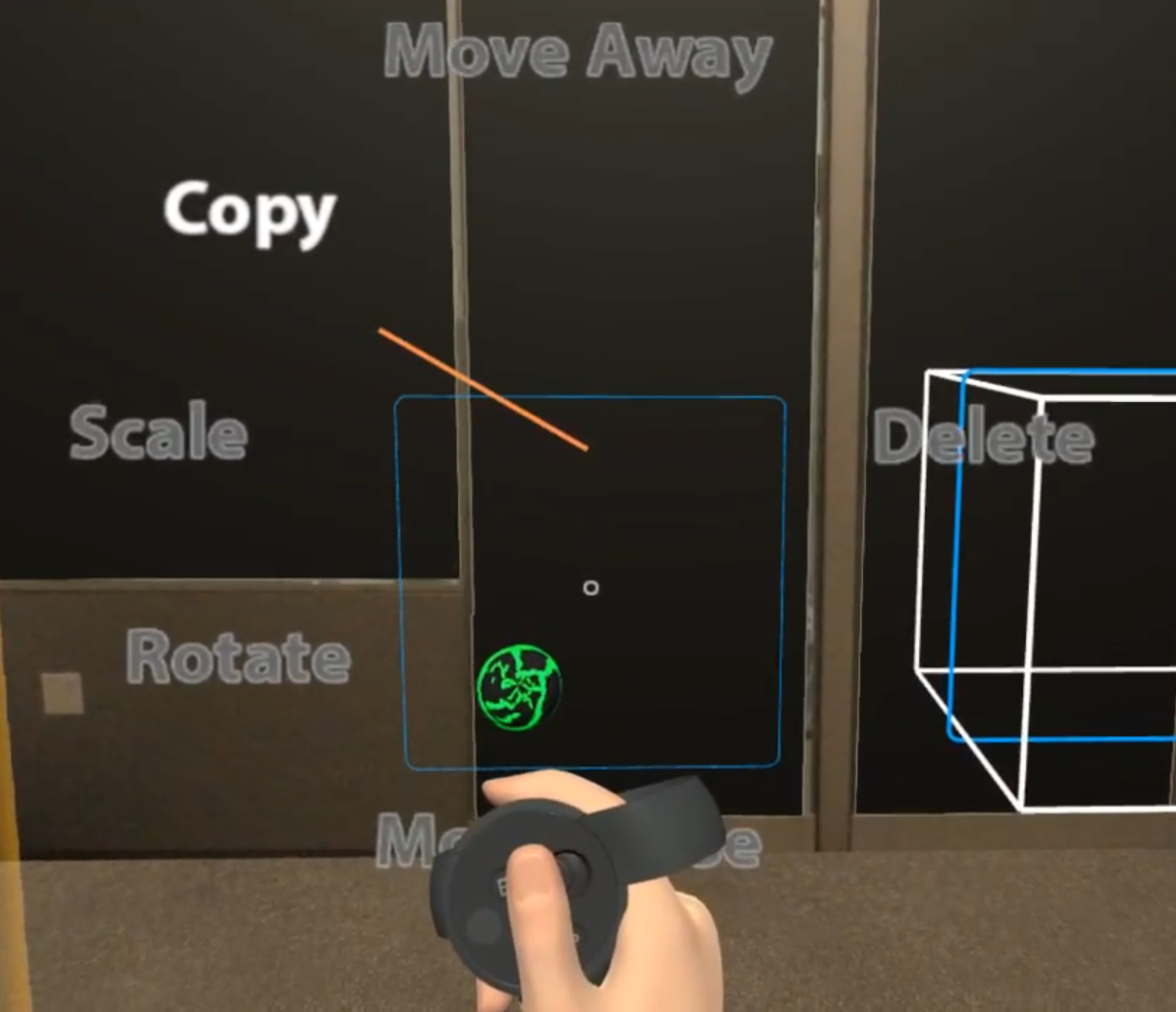}
        \caption{the user has moved the copy bounding box to the desired location}
    \end{subfigure}
    \begin{subfigure}[b]{0.4\columnwidth}
        \includegraphics[width=1\columnwidth]{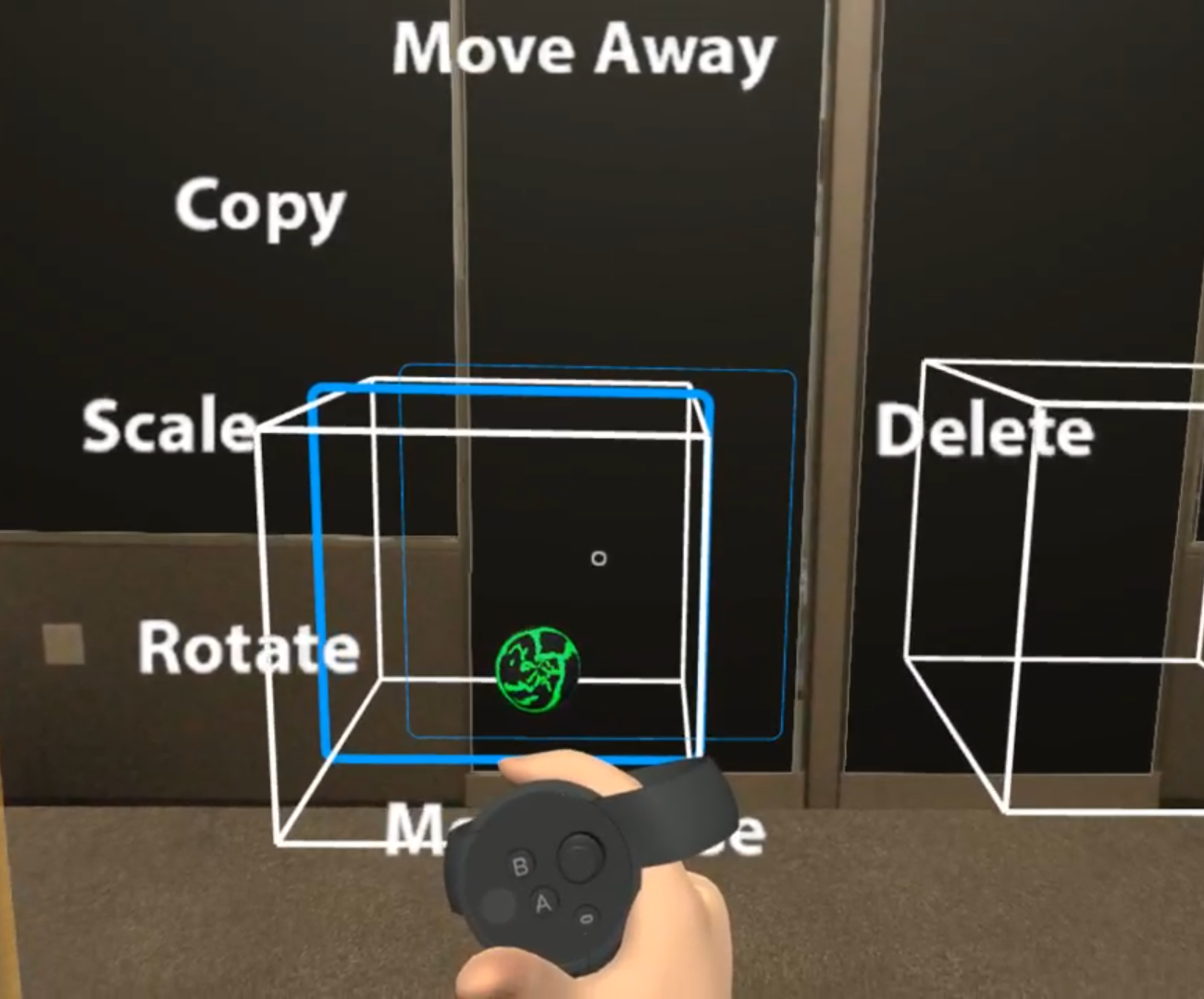}
        \caption{the copy operation is committed, and the cube is copied to the final location.}
    \end{subfigure}
    \caption{Upon selection of a 3DSketch, the user is shown a pie menu which can be used to manipulate the object. The supported operations (going counterclockwise) are deletion, moving the sketch away in 3D, copying the sketch, scaling, rotating and moving the sketch. Operations can be chained together. For example,
    the user can copy a 3DSketch repeatedly or rotate and scale in sequence without deselecting.}
\end{figure}

\section{CollabVR Implementation}
CollabVR is a VR/MR compatible system that supports drawing interactive content and provides a shared-space environment for local or distributed group communication and collaboration. To implement the whole system, we need a content creation server to provide elements to display and interact with. It also requires network frameworks to transmit the data from the content creation server to each client. 
\subsection{Chalktalk} \label{sec:ct}
Chalktalk is a web browser-based 3D presentation and communication tool written in JavaScript, in which the user draws interactive \textit{Sketches} for presentation. It extends the functionality of a slide-show, allowing the user to draw interactive and combinable objects with a mouse.

\subsection{Network}
\begin{figure}[h]
 \centering
 \includegraphics[width=0.75\columnwidth]{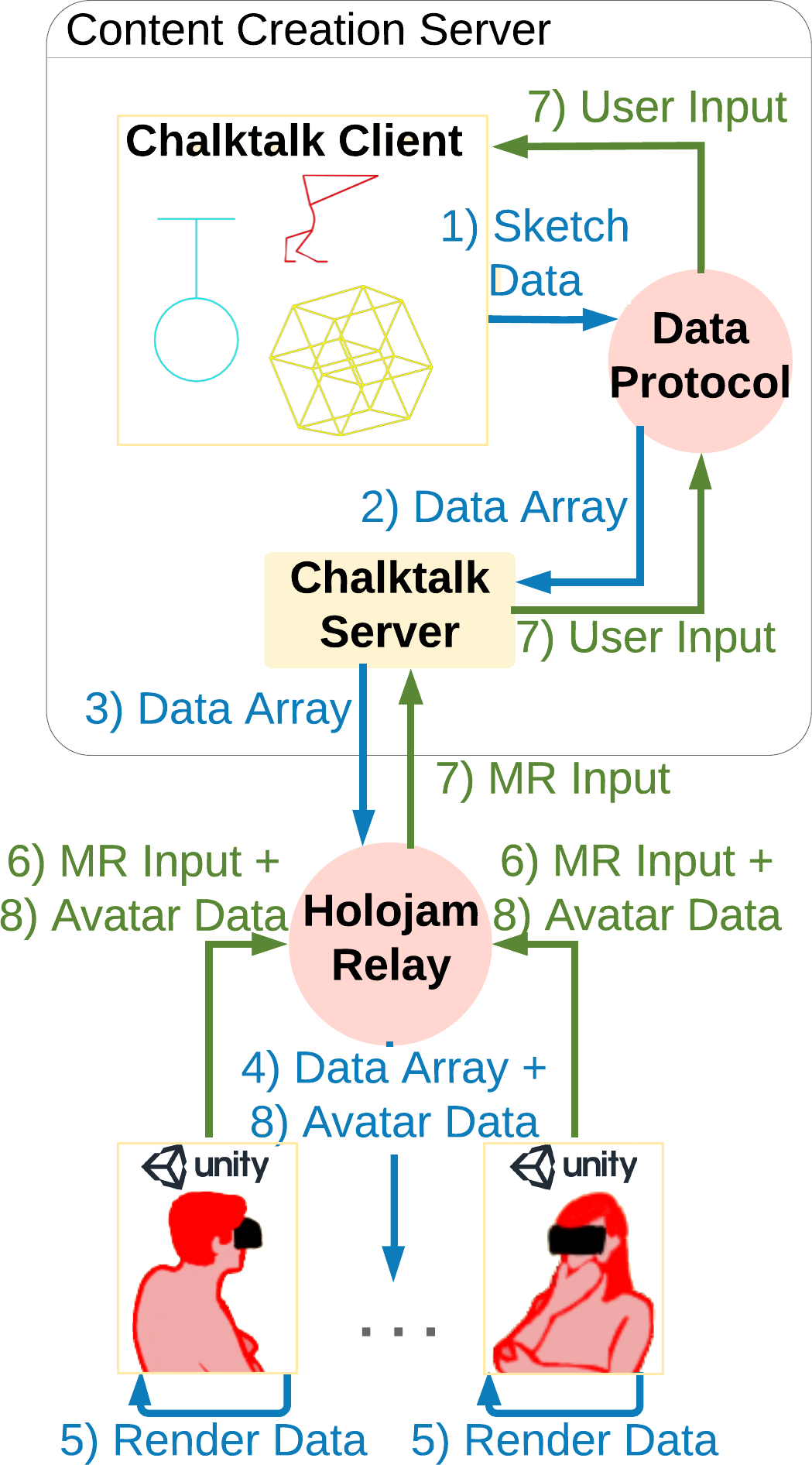}
 \caption{Network Diagram for CollabVR. 
 1) \textit{Sketch} data are serialized on the Chalktalk client side every frame and 2) sent as a data array to the Chalktalk server, 3) to the Holojam relay, and then 4) to all Unity clients, where 5) the data are rendered on content board(s). 6) MR user input is sent back through the pipeline to the Chalktalk client and 7) translated into HTML canvas mouse events. 8) Avatar synchronization data are also sent between clients using the Holojam relay.}~\label{fig:datapipeline}
\end{figure}

We used two network frameworks for this system: Holojam and Holodeck. Holojam~\cite{perlin2016future} is a shared space network framework, written in Node.js and C\#. It synchronizes data across devices and supports custom data formats. For this system, we customize the data protocol to send points and 3D meshes needed for our system. Chalktalk sends the data to the server, which broadcasts to all the clients. The clients send their user-specific data in the reverse direction. (See figure~\ref{fig:datapipeline} for a detailed data pipeline diagram.)
Holodeck is a real-time network framework written in Node.js. We use Holodeck to support audio synchronization.

\section{User Study}

\subsection{Internal telepathy mode Tests}
For simplicity, we decided to use only one of the two versions of telepathy mode for the subsequent user study, and we conducted internal tests to choose. We found that \textit{Immersive Telepathy} caused those less familiar with movement in VR to have a sense of disorientation when viewing the world from others' perspective. Some did not have this feeling. So as not to introduce this additional variability and limit the candidate pool for our user study, we opted for \textit{Windowed Telepathy}. We believe that \textit{Immersive Telepathy} mode requires more development and testing.

\subsection{Main User Study}
We conducted a user evaluation with groups of four people at a time to gain feedback about the usefulness and effectiveness of the system and each configuration. We disabled the ability to switch between configurations so we could run a separate experiment per individual configuration. We are particularly interested in whether different configurations have different impacts on communication and how this system improves communication.

\subsection{Participants}
12 participants (5 female and 1 left-handed) aged from 20 to 30 (mean = 23.58, sd = 3.45) were recruited via email and word-of-mouth to evaluate CollabVR. Most participants were students or staff concentrating in the area of Computer Science. However, not all the participants had rich VR experience. The CollabVR experience and interviews were recorded and all participants were informed. Each participant gave consent.

\subsection{Apparatus}
CollabVR was implemented in Unity Engine on desktop computers with Nvidia GTX 1080 cards. We used Oculus Rift CV1 HMDs with two Oculus Touch controllers, which we paired with one of four computers during the experiments. We connected the four computers through Holojam and Holodeck with Ethernet cables.

\subsection{Procedure}
The user study session contained the following steps:
\subsubsection{Introduction and Training (30-40 minutes)}
Participants were first given an introduction to the user study. Then they were given a 10 minute lecture through a large monitor on how to use Chalktalk. We described how to do freehand drawing, how to create pre-defined 3D objects, and how to change the color of objects.
Then for 10 minutes participants were given a live demo on how to use CollabVR. One experimenter put on the headset and described how to use each controller button and the functionality in the system, including getting permission to draw, manipulating drawings and objects in 3D and using telepathy mode.
After that, each participant was moved to separate virtual rooms. They learned how to use CollabVR individually for 10 minutes. Then they were moved to the same shared virtual room to learn how to observe other people.

\subsubsection{Experiment with Three Configurations (35-45 minutes)}
All the groups were asked to experience three 10-minute sessions. The configuration was changed for each session. We counterbalanced the order of the configurations for each group using Latin Square. For each session, the participants were asked to design a living room containing only three items: one table, one chair and one sofa. The steps were:
\begin{enumerate}
    \item Each participant wrote down which item they picked, the details of the design they expected to create and the placement they envisioned for  the three items
	\item Each joined the VR experience to share what was in their minds using our system. To reduce the learning curve, we drew the entire dictionary of pre-defined 3D objects on one board for them to copy and use as they wanted. Participants could choose to stand or sit.
	\item Each participant exited the VR experience and wrote down their final decisions for the design of the living room.
\end{enumerate}
Because we only provided three items for the group (four people), more than one person was guaranteed to pick the same item. They needed to resolve conflicts and come up with a final decision for the design of living room.

\subsubsection{Questionnaire and Interview (10-15 minutes)}
We collected participants' feedback during the user study with a questionnaire and interview. One configuration-focused questionnaire was provided to participants after each session. The questionnaire contained 7-point Likert scale questions to gather participants' opinions on how they expressed and perceived ideas with the given configuration. One general questionnaire and a group interview was conducted after all the sessions were completed to gather feedback on the three configurations and how well CollabVR helped the participants communicate.

\subsection{Results}

We asked each participant for their opinions on the importance of (1)keeping the content in the view, (2) keeping in the view or being aware of the presence of the person who is drawing (7 = extremely important). For (1), everyone agreed it was more than moderately important to keep the content in the view ($M=6.00$, $SD=0.60$). For (2), 2 of them had a neutral opinion and P11(M) thought it was somewhat important to be aware of the person who is drawing ($M=5.50$, $SD=1.17$). 

\begin{table}
 \centering
 \begin{tabular}{c p{0.85\linewidth}}
    \toprule
    & Questions for Each Configuration\\
    \midrule
    Q1 & How would you rate the configuration?\\
    Q2 & How hard/easy was it to use the configuration?\\
    Q3 & How unhelpful/helpful was the configuration for completing tasks?\\
    Q4 & How hard/easy was it to follow the content being created?\\
    Q5 & How hard/easy was it to follow the project partner who is drawing?\\
    Q6 & How unhelpful/helpful was this configuration for following other people's ideas?\\
    \bottomrule
 \end{tabular}
 \caption{Questions for Each Configuration}~\label{tab:questions}
\end{table}

\begin{table}
 \centering
 \begin{tabular}{c p{0.85\linewidth}}
    \toprule
    & Questions for VR Experience\\
    \midrule
    Q7 & The telepathy mode helped you to have the same idea with your project partners.\\
    Q8 & The tool helped you show/express to your project partners what you were thinking.\\
    Q9 & I would use this tool to collaborate on my own projects.\\
    \bottomrule
 \end{tabular}
 \caption{Questions for VR Experience}~\label{tab:questions}
\end{table}

\begin{table}[h]
 \centering
 \begin{tabular}{c c c c c c c}
    \toprule
     & \multicolumn{2}{c}{Side-by-side} & \multicolumn{2}{c}{Mirrored Face-to-face} & \multicolumn{2}{c}{Eyes-free}\\
     & M & SD & M & SD & M & SD\\
    Q1 & 5.42 &0.99 & 6.08 &0.79 & 4.42 &1.56\\
    Q2 & 5.67 &0.98 & 6.00 &1.04 & 4.00 &1.71\\
    Q3 & 5.17 &1.03 & 6.17 &0.72 & 4.50 &1.38\\
    Q4 & 5.17 &1.34 & 6.00 &1.28 & 4.42 &1.31\\
    Q5 & 5.25 &1.54 & 5.67 &1.44 & 4.33 &1.30\\
    Q6 & 5.42 &1.51 & 5.58 &1.16 & 4.17 &1.34\\
    \bottomrule
 \end{tabular}
 \caption{Results for Each Configuration}~\label{tab:results}
\end{table}

For each configuration, participants were asked about the feeling, the usability, the effectiveness of that configuration with respect to multiple factors. See the full questions and results in table~\ref{tab:questions} and table~\ref{tab:results}. We conducted a repeated measures MANOVA model on the quantitative results. From the statistical results, there is no significant difference (p<0.005 means significant) among different configurations with respect to all the questions. That means all the configurations are acceptable to some extent.

\subsubsection{General Feedback for Each Configuration}

We asked each participant to rank the three configurations after they tried all of them. 7 ranked mirrored face-to-face, 3 ranked side-by-side, and 2 ranked eyes-free as their favorite configuration.

We can see from Q1 to Q3 that participants responded very positively to mirrored face-to-face configuration.
Participants agreed that mirrored face-to-face helped them to focus on the board, both when drawing and watching others. P3(F) commented that participants \textit{``can just communicate with you because I feel that you're just in front of me. I don't need to find where you are''}. 
Participants who liked the mirrored face-to-face configuration most linked the configuration with collaboration and productivity.
\begin{quote}
    The second one, mirror, was my favorite, because it almost felt like we were working on a project together, and everyone's leveled and everyone's working on the same board because you see everyone across from you. I think for that one we made the best stuff.(P9, F)
\end{quote}
P5(M) also said: \textit{``I think I was most productive because I could see both the board and the people and they wouldn't block my view.``}. Furthermore, P10(M) felt that \textit{``In the mirror you can easily see who is drawing and also for communication I think it's a little bit easier, intuitive to know who is talking.``} 
Most participants found mirrored face-to-face created the illusion of a large space and reacted positively to having room to move around. \textit{``When ... facing others [in Mirrored face-to-face] I think my space is kind of free and clean.``}P3(F). One participant(P7, F) felt the physical separation between avatars and the open space made her feel distant from her peers. \textit{``I felt so alone because I was far away from my partners''}.

From side-by-side configuration, participants responded positively with respect to Q1 to Q3. 
P5(M) found side-by-side \textit{``encouraged more conversation because we were next to each other''} even if it is \textit{``a little more cluttered''}.
2 participants (P1, M and P2, M) who ranked side-by-side as their favorite configuration shared a similar opinion that \textit{``the real world is more similar to Side-by-side.''}, and both of them were new to VR (only tried VR for less than 1 hour). The side-by-side configuration seems to be more beginner-friendly than the other two. 

Only 2 participants (P7, F and P10, M) prefer the eyes-free configuration and they responded diverged for the eyes-free configuration. They stated, eyes-free \textit{``allows drawing on the table, which is more intuitive to draw''}(P7, F) and \textit{``I felt like I had more control over what I was drawing''}(P10, M). In their view, eyes-free works better for detailed and long-term drawing.
 P2(M) found it hard to use because of the mapping between the surface board and vertical board, \textit{``It is very hard to track what we are drawing.''}. P7(F) thought it was easy to work on because \textit{``I was using it to draw in front of me as if I were at a desk''}. For P11(M), although he rated this configuration with a lower score, he did foresee potential.
\begin{quote}
    I think this configuration has a great deal of potential if you had haptic feedback or other controller types (a pen and some surface to draw upon). I didn't think it affected my ability to sync up with my partners, but I was trying to remap the axes that my mind expected vs the orientation of the table (P11, M).
\end{quote}
Some participants found it easiest and more enjoyable to look down at the horizontal drawing surface instead of at the vertical board as expected in the eyes-free configuration. P10(M) commented, \textit{``If you look at the table then there's no mapping of the axes. Is the same. At the very beginning I can't draw the circle when I'm looking at the board, but it's very easy when I look at the table.''} This indicates that the eyes-free configuration has a greater learning curve than the others, as users sometimes used the horizontal surface naturally to avoid remapping the axes of the controls. 


\subsubsection{Following the Content in Each Configuration}
From the results (Q4 and Q5), we can see participants responded positively to both mirrored face-to-face and side-by-side configurations, and neutrally to the eyes-free configuration.
P7(F) felt \textit{``their heads and hands blocked my view of the board''} in the side-by-side configuration. In contrast, participants think mirrored face-to-face helped them express the ideas. \textit{``I think this was ideal if you wanted to show someone your idea''}(P11,M). And it helped with understanding ideas too. \textit{``I was able to see them and the drawings at the same time, which was great''} and \textit{``very in sync because no one's avatar was in my way''} (P6,F). Similarly, 
\begin{quote}
    In side-by-side, people block you a lot and it is hard to move around to find a comfortable place to go, and it's also hard to find who's doing what. In the mirror you can easily see who is drawing and also for communication it's a little bit easier, intuitive to know who is talking(P10, M).
\end{quote}
From the results and interview, we can tell that mirrored face-to-face has a better effect on helping follow the content. We noted that people behaved differently depending on whether they chose to stand or sit during the experience. When participants (like  P8, M) sat (i.e. could not move around as much as when they were standing), they seemed to find mirrored face-to-face more helpful for keeping track of the content and the other participants, whereas for side-by-side, 



\subsubsection{Telepathy Mode}
We asked participants for their thoughts on telepathy mode (Q7). They responded somewhat positively to this mode ($M=4.67$, $SD=1.23$). P6(M) pointed out it could be helpful for a presentation scenario, \textit{``Maybe if someone was teaching or talking, and [you could] observe them while they write.''} P8(M) agreed that he \textit{``observed people who were drawing. It is easier to see what they were doing.''} P1(M), P4(M) and P10(M) did not try telepathy mode and P9(F) found it \textit{``distracting.''} 
P12(F) emphasized that telepathy mode would likely make the most sense in larger spaces in which people are more distant from each other.

Some participants found telepathy mode entertaining. P5(M) \textit{``was observing the person next to me and sticking my head in their face.''} and P7(F) found \textit{``it's just funny. I just liked having the window up.''}. telepathy mode is designed to provide more alternatives for seeing the world when the view of the participants or content is blocked. For participants who experienced it for serious reasons, it had positive influence during communication. We anticipated the mode could also be entertaining.

\subsubsection{Communication Effectiveness}

For Q8, participants responded with positive feedback indicating that the system helped them with communication ($M=.75$, $SD=0.87$).
P3(F) described using the system's 3DSketches to communicate to P2(M) what type of chair she wanted to draw. She and P2(M) reached a final decision for the design together:
\begin{quote}
    I think when P2(M) draws the legs [of the chair], I quickly get his idea about the design of the legs, so he doesn't need to say what kind of legs he wants. I just say maybe we can try a round surface, and then we have it. And he shows me the shape .... so we choose a round shape quickly.
\end{quote}

For Q9, we received widely distributed answers (see figure~\ref{}). P2(M) and P10(M) strongly disagreed that they would use the system on their own projects. VR was new to both of them, so the learning curve of the new platform and interactions likely influenced their opinions strongly. Meanwhile, 3 participants agreed and 3 participants somewhat agreed that they would use the system to collaborate on projects with others. P8(M), a digital artist and 3D animator, said:
\begin{quote}
    I'd use this for my own work. It's like a more --not quite as artsy -- business tool. I would use it to get my ideas across to people.
\end{quote}
This suggests that the system may be appealing for sharing ideas as part of professional work.
\subsubsection{Summary and Other Findings}

Participants reflected on the importance of using visual representations of ideas for communication, and generally expressed what they wanted using drawings in addition to verbal communication. For example, P8(M) said \textit{``Just being able to draw it is enough because it's a visual representation of what you want to say.''}. Most people agreed that the system shows promise as a prototyping, ideation, and brainstorming tool that would be used to show concepts and do early designs, or to show things to each other (P3, F, and P2, M and P8 M for example). Participants, such as P8(M) and P11(M) commented on how the system could be used for business and practical applications. Others like P7(F) and P9(F) seemed to base their opinions on how entertaining the system was. P9(F) preferred the configurations in which she could draw on the vertical board, as \textit{``because it's more fun--it's like graffiti almost--cause I could take a tablet and write normally but it's more fun when you get it right on the wall.''}. The overall feedback suggests that the CollabVR system has multiple audiences: those interested in prototyping/drafting or presenting/showing ideas as part of professional content creation work, teaching, or business, as well as those who want to communicate ideas casually and in a fun way with one another.

\section{Limitations and Future Work}
For future work, we first wish to refine the configurations and telepathy mode and to explore how they scale to larger spaces. We intend to try alternative implementations, controls and use cases for eyes-free configuration as well. One possibility for improving how comfortable it is to draw in \textit{Eyes-free} is to have users sit at a mobile desk on which they can draw using a pen-like device and rotate themselves to work at multiple virtual content boards, while using one physical surface. As indicated in the user study, eyes-free seems to have a steeper learning curve than the others, but may have potential given additional development and alterations.
One limitation of our study was the fairly small maximum tracking spaces and mobility constraints of the Oculus Rift device. Recently announced tetherless VR technologies promise greater mobility and larger tracking spaces that would allow for larger groups and environments, which we expect would change the dynamics of communication in our system. Improving telepathy mode is related to exploring larger spaces, in which users believed the mode would find more use. We will go more deeply into alternative designs for telepathy mode, which were not the focus of this paper. We anticipate combining the advantages of the immersive and windowed implementations.
In addition, we are interested in investigating how communication and collaboration are affected when users are permitted to switch configurations at their discretion.

\section{Conclusion}

We have presented CollabVR, a multi-user VR collaboration environment for communication, with three different configurations of participants and content (\textit{3DSketches}). The \textit{configurations} are 1) side-by-side, 2) mirrored face-to-face and 3) eyes-free. We ran a user study to compare the three configurations and gauge users' opinion on them and the system as a whole. User study participants suggested that CollabVR has potential to be a prototyping tool for design work and ideation, as well as for recreation. Side-by-side appealed most to those without prior experience with VR, as it was closest to a whiteboarding environment in real life. The mirrored face-to-face configuration was the favorite configuration--most users felt it facilitated collaboration and productivity, and made it easy to tell who was speaking at a given moment. Users felt that the eyes-free configuration had merit despite its steep learning curve, and preferred to look down at the configuration's horizontal drawing board rather than the vertical board. We are interested in exploring variations of this configuration further. Telepathy mode, which allowed users to see from each other's perspectives, was designed to help people understand ideas and see content that would otherwise be blocked. Based on the study, this mode also has potential to be explored in future iterations.
Overall, our study indicates that CollabVR has the potential to be a useful tool for communication.




\section{Acknowledgments}
We thank all the volunteers for participating in our experiments and video shoots.

%
%
%
%
%
\balance{}

\section{References Format}

\balance{}

\bibliographystyle{SIGCHI-Reference-Format}
\bibliography{sample}

\end{document}